\documentstyle[aps,epsf]{revtex}
\input epsf.sty
\begin{document}
\draft
\flushbottom
\twocolumn[
\hsize\textwidth\columnwidth\hsize\csname @twocolumnfalse\endcsname

\title{Subgap resonances and conduction channels in mesoscopic 
superconducting devices.}
\author{Elena Bascones$^{1,2}$ and Francisco Guinea$^2$}
\address{$^1$ Departamento de F{\'\i}sica de la Materia Condensada. Universidad Aut\'onoma de
Madrid. E-28049 Madrid. Spain \\
$^2$Instituto de Ciencia de Materiales, Consejo Superior de Investigaciones
Cient{\'\i}ficas,
Cantoblanco, E-28049 Madrid, Spain}
\date{\today}
\maketitle
\tightenlines
\widetext

\begin{abstract}
Peaks associated to multiple Andreev reflections have been used
to identify the number of conduction channels, and their
transmitivity, in mesoscopic superconducting junctions.
We analyze the influence of the detailed shape of each channel,
and the number of weak links within it, on the final results.
Connexions with the statistical interpretation of conductance
histograms is also made.
\end{abstract}

]
\narrowtext
\tightenlines

Many experiments show that the conductance of narrow constrictions
between different metals is 
quantized\cite{vW88,WT88,MRJ92,A93,P93,CK95,O98,HL97,KN97,CK97,YR98}. 
The nature of the
structures which give rise to this effect is not well known. 
The forces transmitted through the constriction show
discontinous jumps correlated with the conductance steps\cite{R96},
suggesting that abrupt atomic rearrangements give rise to the
conductance steps. Direct observations show irregular,
and even amorphous structures near the 
contact\cite{A95,PM95,OL94,K95,KY97,KT97,OK98}.

The simplest explanation of the observed quantization describes 
the region near the constriction in terms of
a fixed number of almost fully open channels. 
Such a picture is consistent with models in which the constriction
is described in terms of nearly free 
electrons\cite{B95,BR96,T96,GS97,L97,WM97}.
Related schemes have been used to describe the force oscillations
observed experimentally\cite{SB97,BO98,RD98,YB98}.
Statistical averages 
based on models where the connexion between the channels
within the constriction and the electrodes are allowed to vary randomly,
support this picture\cite{BGS98}. A model of an $n$-channel constriction
coupled to $N$-channel electrodes, shows the desired features,
when $n \ll N$, and the reflection and transmission of electrons from
the electrodes are taken to be random matrices\cite{B98}. 

An alternative picture, based on local properties of the atoms at
the constriction has also been put forward\cite{CLM98}. The conductance is
defined using localized wavefunctions at the atoms within the
constraint. When the constriction narrows to a single atom at some point,
conductances near a single quantum are obtained.
The bands of metals such as Pb or Al
arise, basically, from the hybridization of s and p atomic orbitals,
while those of Na or Au are mainly derived from s orbitals.
Hence, the number of channels through a one atom constriction
in a nearest neighbor tight binding scheme is at least three for
Al or Pb, while it is one for Au. 

These two pictures can be adapted to give similar results for the
total conductance of the constriction, at least when a single
quantum is observed. The properties of superconducting constrictions,
however, are expected to differ. 
The I-V characteristics of
a superconducting constriction described by a single channel with
one barrier are significantly different from those of three or
more channels, even if the total conductance in the normal state is
the same. The Andreev spectrum below the superconducting gap\cite{P94}
has a much richer structure in the second case\cite{S97}. 
Experimental results show, very conclusively, that the
Andreev spectrum in Al and Pb is better fitted by a model of
three or more superconducting channels in parallel, with a
suitably adjusted transmission coefficient\cite{S97}. Current investigations
try to elucidate if a similar distinction can be made in other
quantities which do not depend linearly on the total
conductance, such as the critical current or the noise spectrum.

\begin{figure}
\epsfysize7 cm
\epsfbox{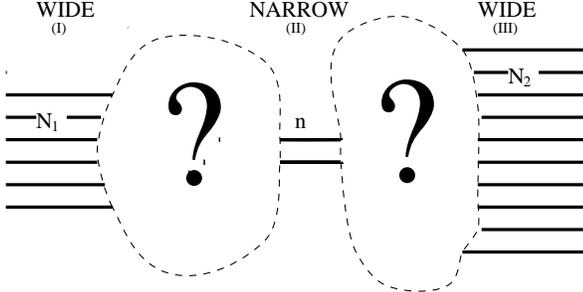}
\caption{Sketch of the representation of the constraint analyzed
in the text. Channels in regions I, II and III are assumed to 
have perfect transmission. The complexity of the junction is represented
by the random matrices connecting the different regions.}
\label{channels}
\end{figure}

We will now focus on the difference between the two models described
above for the case when a single conductance quantum is observed
experimentally. A statistical description based on a nearly free
electron picture assumes that the constriction contains a single
electronic channel. Because of the properties of the (random)
connexions to the bulk electrodes, in most cases, the
transmission of this channel is near one. The larger the number of
channels in the external electrodes, the more likely it is to find
a combination of the channels in the electrodes
with a good matching to the channel in
the constriction, explaining the observed clustering of conductances
near $\frac{2 e^2}{h}$. Making a transformation to a basis
in which these combinations are singled out,
the model can be reduced to a one dimensional single channel problem,
without loss of generality.
The model, however, requires two barriers, at the two connexions
of the weak link with the electrodes, as shown in fig.(\ref{channels}).
On the other hand, a model based on atomic orbitals
with nearest neighbor hybridization, in which the
constraint narrows to a single atom, can be made equivalent,
in the normal state, to a number of electronic channels of the
order of the number of orbitals in one atom. Because of the small
size of the constriction, the description of the transmission through
each of these channels by a single barrier is highly plausible.
Note, however, that the formation of the superconducting condensate
can introduce changes in this picture at scales comparable to
the gap.

In the following, we analyze the I-V characteristics of superconducting
structures made up of a single channel with two barriers, which separate the
constriction from the bulk electrodes.  We can always hybridize the channels
in the bulk  electrodes (regions I and III in figure (\ref{channels}))
in such a way that only one is coupled to the channel within the constriction.
Thus, without loss of generality, the problem is reduced to that of a
single channel interrupted by two barriers. 

We use the scattering formalism
presented in\cite{AB95}, which is equivalent to the Green's function method
formulated in\cite{CML96} and used in the interpretation of
experimental I-V curves . 
Each barrier is defined by a scattering matrix:
\begin{equation}
S_i = \left( \begin{array}{ll} r_i &t_i \\ 
t_i &-\frac{r_i^* t_i}{t_i^*} \end{array}
\right)
\label{matrix}
\end{equation}
where $r$ and $t$ are complex numbers which satisfy $| r |^2 + | t |^2
= 1$, and which are defined up to a global phase. 
In (\ref{matrix}), we assume time reversal symmetry.
The electronic wavefunctions at the channels in the
electrodes and within the constriction, fig.(\ref{channels}), are described as
plane waves, $\Psi_k ( x ) \sim e^{i k x}$. The quasiparticle energies, in the
normal state, are linearized around the Fermi level, 
$\epsilon_k = \epsilon_F + v_F ( k - k_F )$, where $\epsilon_F$
is the Fermi energy, $v_F$ is the Fermi velocity, and $k_F$ is the
Fermi wavevector. Each wavefunction has different phases at the
two barriers. The overall phaseshift includes a contribution from the
region described in terms of perfect channels, and a contribution
from the matching zones, described phenomenologically by the
scattering matrices. The phaseshift can be written 
as an energy independent part
plus an energy dependent contribution, which we parametrize as:
\begin{equation}
\phi ( \epsilon ) = f \frac{\epsilon - \epsilon_F}{\Delta}
\label{phase}
\end{equation}
where $f$ is dimensionless, and $\Delta$ is the superconducting gap.
If the transmission through the constraint is ballistic, 
we have $f = \frac{\Delta L}{\hbar v_F}$, where $L$ is the length
of the region between the barriers. A more general situation is
discussed later.

The model has four adjustable parameters: the absolute values, $| t_i |^2$,
the phase between $r_1$ and $r_2$, which we label $\phi_0$,
and $f$ as defined above.

In the normal state, the system formed by the two barriers in series
has an energy dependent transmission:
\begin{equation}
T ( \epsilon ) = \frac{T_1 T_2}{1 + R_1 R_2 - 2 \sqrt{R_1 R_2} \cos
\left( \phi_0 + 2 f \frac{\epsilon}{\Delta}\right)}
\label{transmission}
\end{equation}
where $R_i = | r_i |^2$ and $T_i = | t_i |^2$ 
are the reflection and transmission coefficients
at each barrier. 

The most general form of the wavefunctions at regions I, II and III, is:


\begin{eqnarray}
\Psi_I^{e} &= &\sum_{m,n} \left[ \left( a_{2n}^{2m} A_n^m 
+ J_0 \delta_{m0}\delta_{n0}
\right) e^{i k x} + B_n^m e^{- i k x} \right] 
\nonumber \\ & &e^{i ( \epsilon + 2 n e V_1 
+ 2 m  e V_2 ) t} \nonumber \\
\Psi_I^{h} &= &\sum_{m,n} \left[ A_n^m e^{i k x} + a_{2n}^{2m} B_n^m
e^{- ikx} \right] e^{-i [\epsilon + 2 n  e V_1 + 2m  e V_2 ] t} \nonumber \\
\Psi_{II}^{e} &= &\sum_{m,n} \left[ ( a_{2n+1}^{2m} \tilde{E}_n^m
+ E_n^m ) e^{ikx} \right. + \nonumber \\ 
& &\left. ( a_{2n+1}^{2m} F_n^m + \tilde{F}_n^m ) e^{-ikx} \right]
e^{-i [ \epsilon + (2n+1) e V_1 + 2 m e V_2 ] t} \nonumber \\
\Psi_{II}^{h} &= &\sum_{m,n} \left[ ( \tilde{E}_{n-1}^m +
a_{2n-1}^{2m} E_{n-1}^m ) e^{ikx} + \right. \nonumber \\
& &\left. ( a_{2n-1}^{2m} \tilde{F}_{n-1}^m
+ F_{n-1}^m ) e^{-ikx} 
\right] e^{- i [ \epsilon + ( 2 n - 1 ) e V_1 + 2 m  e V_2 ) t} \nonumber \\
\Psi_{III}^{e} &= &\sum_{mn} \left[ C_n^m e^{ikx} + a_{2n+1}^{2m+1} D_n^m
e^{-ikx} \right] \nonumber \\
& &e^{-i [\epsilon + ( 2n+1 ) e V_1 + ( 2m+1 ) e V_2 ] t}
\nonumber \\
\Psi_{III}^{h} &= &\sum_{m,n} \left[ a_{2n-1}^{2m-1} C_{n-1}^{m-1} e^{ikx} +
D_{n-1}^{m-1} e^{-ikx} \right] \nonumber \\
& &e^{- i [ \epsilon + ( 2 n - 1 ) e V_1 + 2 m e V_2
] t}
\end{eqnarray}


where $V_1$ and $V_2$ are the voltage drops at the two barriers, $a_n^m (
\epsilon ) = a ( \epsilon + 2 m e V_1 + 2 n e V_2 )$ is the Andreev
reflection amplitude, and $J_0$
in the first equation gives the incoming current. The matching conditions
lead to a set of matrix equations between the coefficients
$A, B, C, D, E, \tilde{E}, F$ and $\tilde{F}$ in the
wavefunctions. It can be shown, after some algebra, that the only finite
coefficients have $m = n$.
This implies that physical quantities depend on the total
voltage drop, $V_1 + V_2$. This is a consequence
of our neglect of inelastic scattering in region II.
Finally, the coefficients in region II,
$E, F, \tilde{E}$ and $\tilde{F}$, can be written in terms of the other four,
leaving a set of equations formally equivalent to those of a single junction. 
They can be solved using recurrent fractions, as described in\cite{BS80}.

The main  difference with the case of a single barrier
is the energy dependence of the phase, parametrized by $f$.
Multiple Andreev scattering is determined by the 
transmission coefficient of the constriction at different energies. 
Hence, it is reasonable to expect that this energy dependence 
has a significant influence on the I-V characteristics below the
gap energy.

\begin{figure}
\epsfysize7 cm
\epsfbox{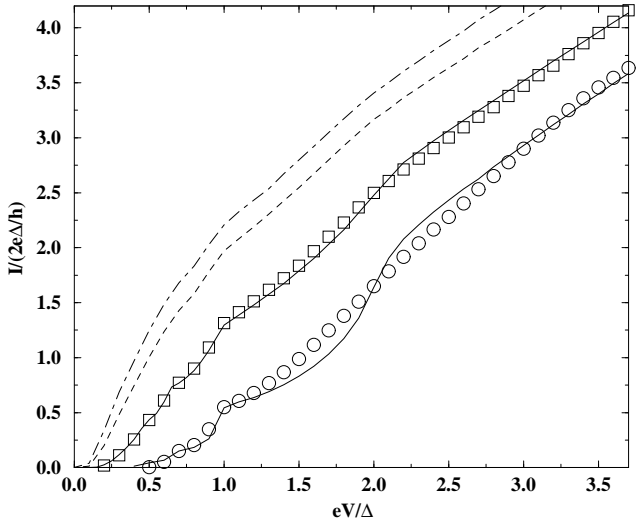}
\caption{I-V characteristics at low voltages of superconducting junctions.
Full curves are obtained with the parameters used in[34]. 
Upper curve: two channels of transmissions $T_1 = 0.74$ and $T_2 = 0.11$.
Lower curve: three channels of transmissions $T_1 = 0.46, T_2 = 0.35$
and $T_3 = 0.07$. Squares and circles are approximations obtained with
the double barrier model described in the text. The broken curves are
results for a single channel, one barrier model with the same normal
state conductance.}
\label{fittings}
\end{figure}

In fig.(\ref{fittings}), we show the best fittings that we obtain to
the I-V characteristics analyzed in (\cite{S97}) in terms of different channels
in parallel. The method of minimization of the mean square deviations
that we use leads to more than one possible fitting of similar quality.
The parameter $f$ is 0.08 for the upper curve (which corresponds
to a two channel
model in\cite{S97}) and 0.1 for the lower curve (which corresponds to a three
channel model in\cite{S97}).

\begin{figure}
\epsfysize7 cm
\epsfbox{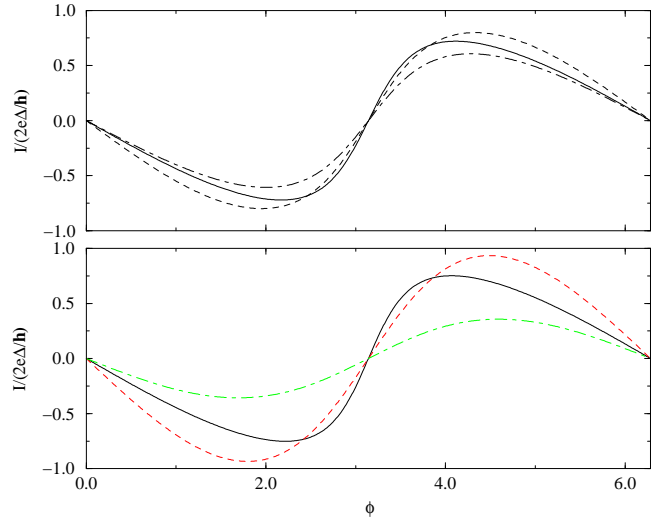}
\caption{Josephson current, as function of the phase difference between
the superconducting electrodes, of the two sets of barriers
described in fig.(2). Full line: single channel with one barrier
and the same transmission coefficient as the set of parallel channels
considered in[34]. Broken line: Results for two or three channels in
parallel. Dash-dotted line: Results for a single channel with two barriers
in series. 
The upper panel corresponds to the two channel case of fig.(2).
The lower panel corresponds to the three channel case.}
\label{current}
\end{figure}

Another quantity which has a non linear dependence on the transmission,
and which can be used to elucidate the detailed structure of the constraint,
is the critical current. In fig.(\ref{current}) we show the dependence
of the Josephson current\cite{B91}
 on the superconducting phase between the electrodes
for the two cases analyzed in fig.(\ref{fittings}) and for a single channel
and one barrier of the same normal state conductance. The maximum
value gives the critical current of the junction. Our results deviate
from the single channel one barrier case in the opposite direction to
those obtained by combining different channels. 

In order to relate $f$ to the structure of the constraint, 
we assume that the motion of
the electrons in the contact zones is diffusive, with a mean free path,
$l$. The dependence of
the phaseshift on distance and frequency goes as $\frac{\omega L^2}{D}$,
where $D \approx \frac{v_F l}{3}$ is the diffussion coefficient.
Replacing $\omega$ by $\Delta$, we obtain:
\begin{equation}
f \sim \frac{\Delta L^2}{\hbar v_F l} \sim \frac{L^2}{\xi_0 l}
\label{phaseshift}
\end{equation}  
where $L$ is the length of the constriction, and $\xi_0$ is the
coherence length at zero temperature. Taking $\xi_0 \sim 10^4$\AA \,
for Al, the values of $f$ that we obtain, $\sim 0.1$, suggest values
of the order of $L \sim 300$ \AA \, and $l \sim 100$ \AA. 
These numbers are consistent
with experimental observations\cite{LD98}. 

The enhancement of the Andreev reflections that we find in a double
barrier geometry is similar to the results obtained for 
transport through a single level between superconducting 
electrodes\cite{LCLM97}.

In summary, we find that existing experiments cannot rule out a description of
many of the weak links which show conductance quantization in terms
of a small number of channels (one for each quantum) randomly coupled
to larger systems described by a much higher number of channels.
This description is consistent with the observed statistical distribution
of the normal state conductance. Our fitting of experimental results
require constrictions of typical dimensions $L \sim 300$ \AA, and 
strong disorder, with mean free paths $l \sim 100$ \AA. The size of the
constriction, and the existence  of strong disorder, are
compatible with existing experiments.

The model analyzed here, however, needs
not apply to all observed cases, as it only attempts to describe 
an average situation. When the constriction size is of the order
of a single atom, it is plausible conducting evanescent
waves (in our language) play some role. If their contribution to the
conductance 
is a significant fraction of the total value,
a description based on extended wavefunctions
is almost indistinguishable from one based on localized
atomic orbitals. On the other hand, the existence of well defined
conductance steps near integer multiples of $\frac{e^2}{h}$ seems to
us more consistent with a dominance of situations with 
almost fully conducting channels within the constriction. 

We are thankful to P. de Andr\'es, N. Agra{\"\i}t, D. Baereswyl, J.
Costa-Kr\"amer, N. Garc{\'\i}a, G. G\'omez-Santos, A. Levy-Yeyati,
J. G. Rodrigo, G. Rubio, J. J. S\'aenz, P. Serena and S. Vieira for
helpful discussions.
We acknowledge financial support from grant PB96-0875 (Spain). One of us
(E. B.) is grateful to the Universidad Aut\'onoma de Madrid and to
the Comunidad Aut\'onoma de Madrid for a scholarship.

\end{document}